\documentclass[%
 reprint,
 amsmath,amssymb,
 aps,
pra,
]{revtex4-1}
\usepackage{graphicx}
\usepackage{dcolumn}
\usepackage{bm}
\usepackage{verbatim}
\usepackage{makecell}
\usepackage{amsmath,amssymb,amsfonts}

\usepackage[dvipdfm,colorlinks,linkcolor=blue, urlcolor=blue, anchorcolor=blue, citecolor=blue]{hyperref}
\begin{document}

\title{Cavity-Heisenberg spin chain quantum battery}

\author{Fu-Quan Dou}
\email{doufq@nwnu.edu.cn}
\affiliation{College of Physics and Electronic Engineering, Northwest Normal University, Lanzhou, 730070, China}
\author{Hang Zhou}
\affiliation{College of Physics and Electronic Engineering, Northwest Normal University, Lanzhou, 730070, China}
\author{Jian-An Sun}
\affiliation{College of Physics and Electronic Engineering, Northwest Normal University, Lanzhou, 730070, China}


\begin{abstract}
We propose a cavity-Heisenberg spin chain (CHS) quantum battery (QB) with the long-range interactions and investigate its charging process. 
The performance of the CHS QB is substantially improved compared to the Heisenberg spin chain (HS) QB. When the number of spins $N \gg 1$, the quantum advantage $\alpha$ of the QB's maximum charging power can be obtained, which approximately satisfies a superlinear scaling relation $P_{max} \propto N^{\alpha}$. For the CHS QB, $\alpha$ can reach and even exceed $1.5$, while the HS QB can only reach about $\alpha=0.75$. We find that the maximum stored energy of the CHS QB has a critical phenomenon. By analyzing the Wigner function, von Neumann entropy, and logarithmic negativity, we demonstrate that entanglement can be a necessary ingredient for QB to store more energy, but not sufficient.
\end{abstract}

\maketitle


\section{INTRODUCTION} \label{section1}
With the development of quantum science, the potential usefulness of quantum technology in the energy field has attracted a considerable number of authors to introduce and study ``quantum battery (QB)'' i.e., a quantum system that stores or supplies energy \cite{PhysRevE.87.042123,PhysRevLett.118.150601,Bhattacharjee2021,Niedenzu2018,Giorgi2015}. Different from the traditional batteries, the QB usually refers to devices that utilize quantum degrees of freedom to store and transfer energy based on quantum thermodynamics \cite{PhysRevApplied.14.024092,Skrzypczyk2014}. Up to now, considerable attention has been mostly focused on the charging process including the QB's work-extraction capabilities \cite{campaioli2018quantum,shi2022entanglement,Chen2020,Rosa2020,PhysRevLett.124.130601,PhysRevLett.125.236402,Binder_2015,PhysRevE.94.052122,PhysRevA.104.L030402,PhysRevLett.111.240401,PhysRevLett.118.150601,PhysRevLett.122.047702,Friis2018precisionwork,PhysRevE.102.052109,PhysRevResearch.2.023095,PhysRevLett.125.040601,Dou2020}, stable charging \cite{Chen2020,Rosa2020,PhysRevE.100.032107,PhysRevLett.124.130601,Dou2022,PhysRevE.101.062114,Kamin_2020}, self-discharging \cite{PhysRevE.103.042118,Dou2020} and dissipation charging \cite{Rosa2020,PhysRevE.101.062114,Mitchison2021chargingquantum,Kamin_2020,PhysRevResearch.2.033413,PhysRevA.102.052223}. Alicki and Fannes suggested that ``entangling unitary controls'', i.e., unitary operations acting globally on the state of the $N$ quantum cells, lead to better work extraction capabilities from the QB, when compared to unitary operations acting on each quantum cell separately \cite{PhysRevE.87.042123}. Further research uncovered that entanglement generation benefits the speedup of work extraction \cite{PhysRevLett.111.240401}. Later on, two types of charging schemes, ``parallel'' and ``collective'' schemes were proposed \cite{PhysRevA.101.032115,PhysRevX.5.031044}. During the charging procedure of a QB, there is a ``quantum advantage'' in the collective charging scheme, that is, when $N \geq 2$, the charging power of the QB is greater than that of the parallel scheme \cite{Chen2020,PhysRevE.99.052106,PhysRevA.97.022106,PhysRevResearch.2.023113,PhysRevB.99.205437,Chang2021,PhysRevE.103.042118,PhysRevA.104.L030402,PhysRevA.103.052220,PhysRevA.100.043833,kim2021quantum,PhysRevA.105.L010201}.

In the quest for such a quantum advantage and potential experimental implementations of QB, various models have been proposed, which can be mainly divided into two categories: quantum cavity model, where arrays of $N$ qubits are coupled to a cavity field \cite{campaioli2018quantum,Binder_2015,PhysRevE.94.052122,PhysRevLett.120.117702,PhysRevB.98.205423,PhysRevLett.122.047702,PhysRevB.99.035421,zhang2018enhanced,PhysRevB.105.115405}, and atomic models (two-level atoms, three-level atoms, spin, etc.) \cite{campaioli2018quantum,shi2022entanglement,PhysRevB.104.245418,Chen2020,Rosa2020,PhysRevLett.124.130601,PhysRevLett.125.236402,Binder_2015,PhysRevE.99.052106,PhysRevB.99.205437,PhysRevResearch.2.023113,PhysRevA.101.032115,PhysRevE.104.054117,PhysRevE.104.024129,PhysRevA.105.022628,PhysRevB.100.115142,PhysRevE.102.052109,PhysRevE.105.054115,PhysRevA.97.022106,PhysRevA.101.032115,PhysRevResearch.4.013172,PhysRevA.103.033715}.

The Heisenberg spin chain (HS) model is a statistical mechanical model of spin systems, which plays a crucial role in accounting for the magnetic and thermodynamic natures of many-body systems \cite{PhysRevA.70.052322,PhysRevA.71.012307,PhysRevB.79.245414,PhysRevLett.96.247203,PhysRevLett.96.257202,PhysRevLett.102.037203,PhysRevA.80.012323,PhysRevA.81.022303}. In recent years, a wide range of work has been done based on HS and discovered rich phenomena, including the effects of anisotropy parameters, the role of boundary conditions \cite{PhysRevLett.99.027205,PhysRevB.79.214408,PhysRevLett.96.247203,PhysRevLett.91.037205,PhysRevA.70.022313,PhysRevA.71.022315,PhysRevA.72.022345,PhysRevA.71.022308,PhysRevA.72.034302,PhysRevA.72.052101,PhysRevA.73.022339,PhysRevLett.96.187201,PhysRevA.77.042309,PhysRevA.77.062305,PhysRevA.79.052109}.
In the field of QB, the HS has also received considerable attention \cite{shi2022entanglement,PhysRevE.104.054117,PhysRevE.104.024129,PhysRevA.105.022628,PhysRevB.100.115142,PhysRevE.102.052109,PhysRevE.105.054115,PhysRevA.97.022106,PhysRevA.101.032115,PhysRevResearch.4.013172,PhysRevA.103.033715}. The spin-spin interactions can yield advantage in charging power over the noninteracting case, and this advantage can grow super extensively when the interactions are long-ranged \cite{PhysRevA.97.022106}. The study on dynamics of the HS QB has shown that the defects or impurities can create a larger amount of quenched averaged power in the QB in comparison with the situation where the initial state is prepared without disorder \cite{PhysRevA.101.032115}. Furthermore, with the proper tuning of system parameters in the HS QB, an initial state prepared at finite temperature can generate higher power in the QB than that obtained at zero temperature \cite{PhysRevA.101.032115}. Novel finding---After adjusting the magnetic field in the charging, interacting rotation-time symmetric chargers have the potential to produce higher charges in QB than corresponding Hermitian chargers \cite{konar2022quantum}.

However, previous work on spin chain model QB focused on nearest-neighbor interactions \cite{shi2022entanglement,PhysRevE.104.054117,PhysRevE.104.024129,PhysRevA.105.022628,PhysRevB.100.115142,PhysRevE.102.052109,PhysRevE.105.054115,PhysRevA.97.022106,PhysRevA.101.032115,PhysRevResearch.4.013172,PhysRevA.103.033715}.
It is well known that the generation of many interesting and exotic physical phenomena relies on the long-range interactions between spins \cite{PhysRevA.81.022303}. On the other hand, it has been verified that an $N$-spins chain coupled to a cavity field can notably enhance the charging power of QB \cite{PhysRevLett.120.117702,PhysRevB.99.205437}. Therefore, the following questions naturally arise: Under the long-range interactions, how does the cavity-Heisenberg spin chain (CHS) QB perform compared to the HS QB?

In this work, we propose a CHS QB with long-range interactions and investigate the performance of the CHS QB (including, as well, the HS QB as a comparison). Here the battery consists of $N$ spins displayed in a collective mode during the charging process, and the charger includes the cavity-spin coupling and the spin-spin interaction. We investigate the effect of the cavity on QB's performance. We are concerned with the dependence of the stored energy and the average charging power of the QB on the spin-spin interaction and anisotropy. In addition, we analyze how the number $N$ of the spins influences the maximum stored energy and the maximum charging power. We are also concerned with the quantum advantage of the maximum charging power of the QB. Finally, we show a critical behavior for the maximum stored energy of CHS QB and introduce the quantum phase transition, Wigner function, von Neumann entropy, and the logarithmic negativity to analyze the critical behavior.

This paper is organized as follows. Sec. \ref{section2} introduces the concept of the CHS QB and the measure to quantify QB's performance. In Sec. \ref{section3}, we give detailed results and analysis of the CHS QB. The critical behavior is dealt with 
 in Sec. \ref{section4}. Finally, a brief summary is given in Sec. \ref{section5}.

\section{QUANTUM SPIN MODEL AS BATTERY} \label{section2}
\begin{figure}[htp]
\centering
\includegraphics[width=0.5\textwidth]{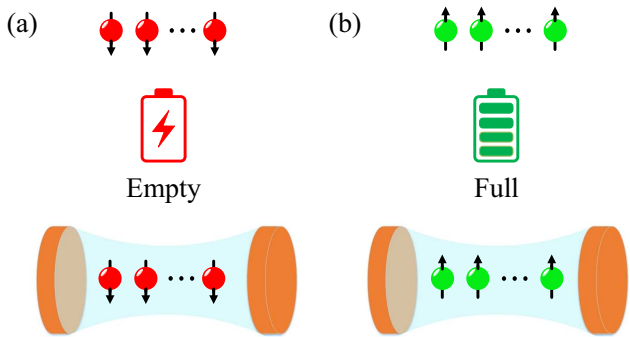}
\caption{\label{fig:model}A sketch of the HS QB and CHS QB. (a) The state of a HS QB and CHS QB when it has no energy to out-put. All the spins are in the ground state, and the battery is empty. (b) The HS QB and CHS QB is in a fully charged state. All identical spins are all in the spin-up state.}
\end{figure}

The CHS QB model consists of a single-mode cavity and HS, which are coupled via the exchanges of photons.  As shown in Fig.~\ref{fig:model}, (a) and (b) are the initial and charged states of HS QB and CHS QB, repectively. The Hamiltonian of CHS QB can be written as (while the Hamiltonian of HS QB is shown in Appendix~\ref{appendix1}. Hereafter, we set $\hbar = 1$)
\begin{eqnarray}\label{HamQB}
H(t)=H_{B}+\lambda(t)H_{C},
\end{eqnarray}
where
\begin{eqnarray}\label{HamB}
H_{B}=\omega_{a}\hat{J}_z,
\end{eqnarray}
\begin{eqnarray}\label{HamC1}
\begin{split}
H_{C}=&\omega_{c}\hat{a}^{\dagger} \hat{a} + g_{1}\sum_{i=1}^{N}\hat{\sigma}_{i}^{x}\left(\hat{a}^{\dagger} +\hat{a}\right)\\
&+\omega_{a}g_{2}\sum_{i<j}^{N}[(1+\gamma)\hat{\sigma}_{i}^{x}\hat{\sigma}_{j}^{x}+(1-\gamma)\hat{\sigma}_{i}^{y}\hat{\sigma}_{j}^{y}+\Delta\hat{\sigma}_{i}^{z}\hat{\sigma}_{j}^{z}].\\
\end{split}
\end{eqnarray}
Here the time-dependent parameter $\lambda(t)$ describes the charging time interval, which we assume to be given by a step function equal to $1$ for $t\in[0,T]$ and zero elsewhere. $\hat{\sigma}_{i}^{\alpha}$ is the Pauli operators of the $i-$th site. $\hat{J}_{\alpha}=(\hbar/2)\sum_{i}^{N}\hat{\sigma}_{i}^{\alpha}$ with $\alpha = x,y,z$. $\hat{a}$ $(\hat{a}^{\dagger})$ annihilates (creates) a cavity photon with the cavity field frequency $\omega_{c}$ and the strength of the spin-cavity coupling is given by the dimensionless parameter $g_{1}$, the $\omega_{a}$ is the frequency of spins. $\gamma$ and $\Delta$ are the anisotropy coefficients and $N$ is the number of spins. We focus on the resonance regime, i.e., $\omega_{a}=\omega_{c}=1$, to ensure the maximum energy transfer. Off-resonance case $\omega_{a} \neq \omega_{c}$ will not be discussed since they are characterized by a less efficient energy transfer between the cavity and spins.

By introducing the ladder operator $\hat{J}_{\pm}=\hat{J}_{x}{\pm}i\hat{J}_{y}$  
 the Hamiltonian $H_{C}$ can be rewritten as
\begin{eqnarray}\label{HamC2}
\begin{split}
H_{C}=&\omega_{c}\hat{a}^{\dagger} a + 2g_{1}\hat{J}_{x}\left(\hat{a}^{\dagger} +a\right)\\
&+\omega_{a}g_{2}[\hat{J}_{+}\hat{J}_{-}+\hat{J}_{-}\hat{J}_{+}+\gamma(\hat{J}_{+}^{2}+\hat{J}_{-}^{2})\\
&+2 \Delta \hat{J}_{z}^{2}-\frac{N}{2}(2+\Delta)].\\
\end{split}
\end{eqnarray}

By selecting different anisotropy coefficients $\gamma$ and $\Delta$, the Heisenberg model can be further subdivided into the following categories, and QB based on these models have been studied as shown in Table~\ref{TAB.1}.
\begin{table}[htbp]
\begin{center}\
  \setlength{\abovecaptionskip}{0pt}
  \setlength{\belowcaptionskip}{10pt}
     \caption{Heisenberg model classification.}
\label{TAB.1}
\setlength{\tabcolsep}{4mm}{
\begin{tabular}{clc}
\hline\hline
\makecell{$\gamma$} & \makecell{$\Delta$} & \makecell{model}\\
\hline
\makecell[l]{$\gamma={\pm}1$} & \makecell[l]{$\Delta=0$} & \makecell[l]{Ising model \cite{PhysRevB.100.115142,PhysRevA.97.022106}}\\
\makecell[l]{$\gamma=0$} & \makecell[l]{$\Delta=0$} & \makecell[l]{XX model \cite{PhysRevE.104.054117,PhysRevE.105.054115,PhysRevA.105.022628,konar2022quantum}}\\
\makecell[l]{$\gamma=0$} & \makecell[l]{$\Delta \neq 0$} & \makecell[l]{XXX model \cite{PhysRevA.97.022106}}\\
\makecell[l]{$\gamma=0$} & \makecell[l]{$\Delta=1$} & \makecell[l]{XXZ model \cite{shi2022entanglement,PhysRevB.100.115142,PhysRevA.97.022106,PhysRevE.102.052109,konar2022quantum}}\\
\makecell[l]{$0 < |\gamma| < 1$} & \makecell[l]{$\Delta=0$} & \makecell[l]{XY  model \cite{PhysRevA.101.032115,PhysRevA.105.022628,PhysRevE.104.024129,PhysRevA.105.022628}}\\
\makecell[l]{$0 < |\gamma| < 1$} & \makecell[l]{$\Delta \neq 0$} & \makecell[l]{XYZ model \cite{PhysRevA.101.032115}}\\
\hline\hline
\end{tabular}}
\end{center}
\end{table}

In the usual case, Heisenberg QB model is charged by an external driving field \cite{shi2022entanglement,PhysRevB.100.115142,PhysRevA.97.022106,PhysRevE.104.054117,PhysRevE.105.054115,PhysRevA.105.022628,PhysRevE.102.052109,PhysRevA.101.032115,PhysRevE.104.024129}. An energy-charged cavity field in an excited energy state can save as much as an external driving field \cite{PhysRevLett.120.117702}. However, the effect of the coupling between the spin chain and the cavity on the HS QB has seldom been taken into consideration. Therefore, in our QB model, the XYZ HS is the battery part coupled with a cavity field that transfers energy to charge the QB.

In our charging protocol, QB will start charging when the classical parameter $\lambda(t)$ is nonzero. The wave function of the system evolves with time, i.e.,
\begin{eqnarray}\label{psit}
|\varPsi(t)\rangle=e^{-iHt/\hbar}|\varPsi(0)\rangle,
\end{eqnarray}
where the $\varPsi(0)\rangle$ is the initial state of the entire system. We consider the charging process of the CHS QB in a closed quantum system. Here, the $N$ spins are prepared in ground state $|g\rangle$ and coupled to a single-mode cavity in the $N$ photons Fock-state $|N\rangle$. Thus, the initial state of the Eq.~(\ref{HamQB}) is
\begin{equation}\label{CHSpsi0}
|\varPsi(0)\rangle=|N\rangle\otimes|\underbrace{g,g,...,g}_{N}\rangle.
\end{equation}

At a particular time instant $t$ , the total stored energy by the battery can be defined as
\begin{equation}\label{Et}
E(t)=\langle{\varPsi(t)}|H_{B}|\varPsi(t)\rangle-\langle{\varPsi(0)}|H_{B}|\varPsi(0)\rangle.
\end{equation}

The corresponding average power for a given time $t$ can be written as $P(t)=E(t)/t$. To maximize the extractable power, it is important to choose a proper time when the evolution should be stopped. Towards this objective, the maximum stored energy $E_{max}$ (at time $t_{E}$) obtained from a given battery can be quantified as
\begin{equation}\label{Emax}
E_{max}\equiv \mathop{max}\limits_{t}[E(t)]=E[(t_{E})],
\end{equation}
and accordingly the maximum power $P_{max}$ (at time $t_{P}$) reads
\begin{equation}\label{Pmax}
P_{max}\equiv \mathop{max}\limits_{t}[P(t)]= P[(t_{P})].
\end{equation}
A convenient basis set for representing the Hamiltonian is $|n,j,m\rangle$, where $n$ indicates the number of photons. With this notation, the initial state in Eq.~(\ref{HamB}) reads $|\varPsi(0)\rangle=|N,N/2,-N/2\rangle$.

The matrix elements of the Hamiltonian $H$ can be evaluated over the basis set $|n,j,m\rangle$ using the following relations for ladder operator of photons and pseudo-spin \cite{Miguel2011,PhysRevA.85.053831,PhysRevE.67.066203}:
\begin{eqnarray}\label{adag}
\begin{split}
&\hat{a}^{\dagger}|n,j,m\rangle=\sqrt{n+1}|n+1,j,m\rangle,\\
&\hat{a}|n,j,m\rangle=\sqrt{n}|n-1,j,m\rangle,\\
&\hat{J}_{\pm}|n,j,m\rangle = \sqrt{j(j+1)-m(m\pm1)}|n,j,m\pm1\rangle,
\end{split}
\end{eqnarray}
while the matrix elements of the CHS Hamiltonian can be found in Appendix \ref{appendix2}.

We remark that the number of photons is not conserved by the CHS Hamiltonian. It is also not bounded from above; thus, it may take an arbitrarily large integer value. In practice, we need to introduce a cutoff $N_{ph}>N$ on the maximum number $N_{ph}$ of photons within our finite-size numerical diagonalization. This choice allows us to select a case scenario of large $N$ values to calculate the stored energy without making any significant difference. In this paper, we selecting the maximum number of photons as $N_{ph}=4N$. Part of the calculations are coded in PYTHON using the QUTIP library \cite{JOHANSSON20131234}.
\section{QB'S ENERGY AND CHARGING POWER} \label{section3}
In this section, we discuss the charging property of the CHS QB and give a part of the calculation results of HS QB, while we refer the reader to the Appendix~\ref{appendix1} for details.

\begin{figure}[!ht]
\centering
\includegraphics[width=0.4\textwidth]{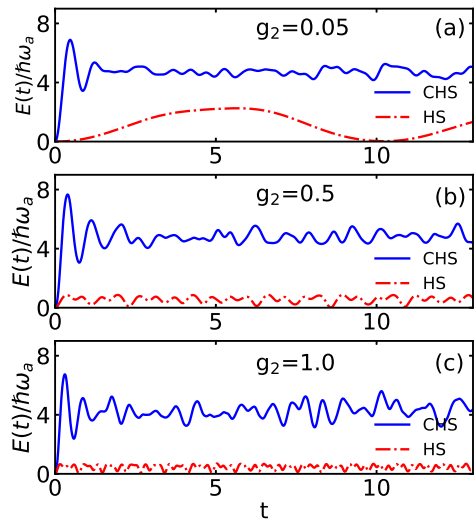}
\caption{\label{fig:Timing}(a) The dependence of the stored energy $E(t)$ (in units of $\hbar\omega_{a}$) on $t$ for different interaction strengths: (a) $g_{2} = 0.05$, (b) $g_{2} = 0.5$ , and (c) $g_{2} = 1.0$.  The CHS QB (blue sold line) and the HS QB (red dash-dot line), respectively. In the paper, all plots are under the same setting of $N = 10$, $g_{1}=2$, $\gamma=0.5$ and $\Delta=2$, unless mentioned otherwise.}
\end{figure}

In order to analyze the effect of the cavity on CHS QB, we illustrate the time evolution of energy as shown in Fig.~\ref{fig:Timing}. It demonstrates that the CHS QB has better performance than the HS QB. In particular, the CHS QB requires less time to achieve the maximum stored energy because of the presence of the cavity.
\begin{figure}[!ht]
\centering
\includegraphics[width=0.5\textwidth]{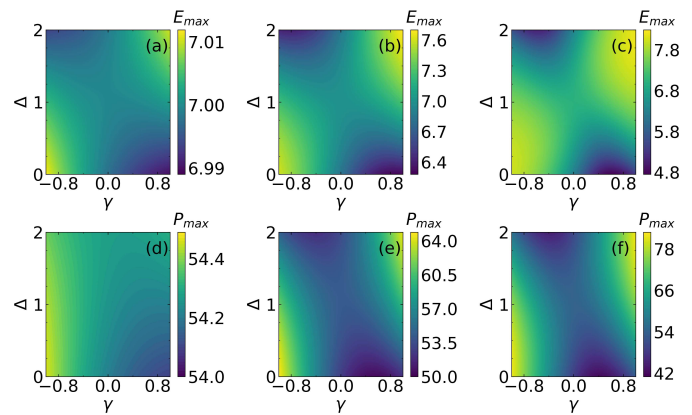}
\caption{\label{fig:deltagamma1}The contour plots of the CHS QB's maximum stored energy $E_{max}$ (in units of $\hbar\omega_{a}$) for (a)-(c), and maximum charging power $P_{max}$ (in units of $\hbar\omega_{a}^{2}$) for (d)-(f) with different values of $g_{2}$. The values of $g_{2}$ are: (a) and (d): $g_{2} = 0.05$, (b) and (e): $g_{2} = 0.5$, (c) and (f): $g_{2} = 1$, respectively. Here, we consider the ranges $\gamma\in[-1,1]$, $\Delta\in[0,2]$.}
\end{figure}

We then calculate the maximum stored energy and the maximum charging power of the CHS QB as a function of the anisotropy coefficients and the results are shown in Fig.~\ref{fig:deltagamma1}. The effect of anisotropy on the maximum stored energy and the maximum charging power of the CHS QB is almost negligible for weak spin-spin interaction strength. This effect becomes more prominent with the enhancement of spin-spin interactions. In other words, stronger spin-spin interaction strength can improve the performance of CHS QB, but also reduces its robustness to anisotropy coefficients. Furthermore, stronger anisotropy can lead to better performance of CHS QB, i.e., when the exchange interactions in the $z$ and $x$ directions are strong ($\Delta\rightarrow 2$ and $\gamma\rightarrow 1$) or the exchange intensity in the $y$ is strong and $z$ direction is weak ($\Delta\rightarrow 0$ and $\gamma\rightarrow -1$), the CHS QB can obtain greater maximum stored energy and maximum charging power. 

The calculation results of both Dicke QB \cite{PhysRevLett.120.117702} and extended Dicke QB \cite{PhysRevB.105.115405} show that for large $N$, the average charging power scales like $P_{max} \propto N^{3/2}$ in single-photon Dicke QB. Therefore, we also expect the existence of a general scaling relation between the charging power of the CHS (or HS) QB and the number $N$ of spins. We assume that the maximum charging power takes the following form
\begin{equation}\label{Pmaxalpha}
P_{max} \propto \beta N^{\alpha}.
\end{equation}
By taking the logarithm, we use linear fitting to obtain the scaling exponent $\alpha$
\begin{equation}\label{logPmax}
log(P_{max})=\alpha log(N)+log(\beta).
\end{equation}
The scaling exponent $\alpha$ essentially reflects the collective nature of the battery in transferring energy.
\begin{figure}[!ht]
\centering
\includegraphics[width=0.4\textwidth]{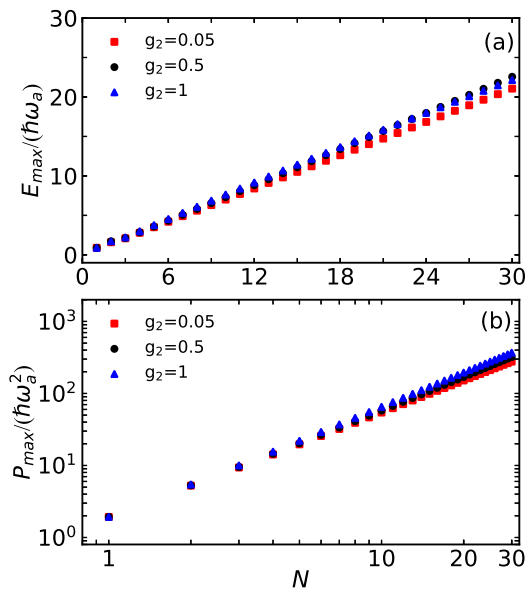}
\caption{\label{fig:variesN2}(a) The maximum stored energy $E_{max}$ (in units of $\hbar\omega_{a}$) as a function of $N$ for different values of  $g_{2}$. Results for CHS QB refer to $g_{2}=0.05$ (red circles), $g_{2} =0.5$ (blue triangles), and  $g_{2} = 1.0$ (red circles). (b) The maximum charging power $P_{max}$ (in units of $\hbar\omega_{a}^{2}$) as a function of $N$. Color coding and labeling are the same as in (a).}
\end{figure}

We find that maximum stored energy and maximum charging power have a clear correlation with $N$ as it increases in Fig.~\ref{fig:variesN2}, which implies that in CHS QB, for large but finite $N$, $E_{max}$ and $P_{max}$ follow the scalling laws:
\begin{equation}\label{EmaxN}
E_{max} \propto N,
\end{equation}
and
\begin{equation}\label{Pmaxalpha1.5}
P_{max} \propto N^{3/2},
\end{equation}
this reveals the quantum advantage $\alpha=1.5$ and even more than $1.5$ of CHS QB can be reached, higher than that of HS QB (scaling exponent of the maximum charging power can only reach $\alpha=0.75$, see Appendix \ref{appendix1}). Similar to the extended Dicke quantum battery \cite{PhysRevB.105.115405}, the scaling rate can be even higher by adjusting the parameters appropriately.
\section{ENERGY CRITICAL BEHAVIOR AND ENTANGLEMENT} \label{section4}
\begin{figure}[!ht]
\centering
\includegraphics[width=0.5\textwidth]{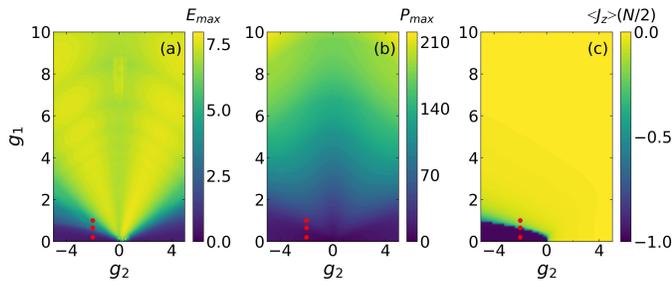}
\caption{\label{fig:contour}(a) and (b) contour plots of CHS QB's maximum stored energy $E_{max}$ (in units of $\hbar\omega_{a}$) and maximum charging power $P_{max}$ (in units of $\hbar\omega_{a}^{2}$) as functions of the cavity-spin coupling strength $g_{1}$ and spin-spin interaction strength $g_{2}$. (c) Phase diagram described by $\left\langle J_{z}\right\rangle/(N/2)$ as functions of the coupling strength $g_{1}$ and interaction strength $g_{2}$. The red dots are three randomly selected points near the critical point.}
\end{figure}

With the consideration of cavity-spin coupling, a quite natural question follows as to the effects on the charging battery. To do so, we calculated the contour maps of $E_{max}$ and $P_{max}$ as a function of the cavity-spin coupling strength $g_{1}$ and the spin-spin interaction strength $g_{2}$, as shown in Fig.~\ref{fig:contour}. In regions of weak coupling strength, the spin-spin interaction strength can significantly affect the maximum stored energy of the CHS QB. However, in regions of strong coupling strength, this effect is almost negligible. Particularly, we find that the maximum stored energy of the QB has a critical behavior, i.e., the system exists a critical point and the maximum stored energy of the QB changes obviously near the critical point.

To clarify such critical behavior, we introduce the quantum phase transition \cite{PhysRevA.78.023634,PhysRevE.67.066203,Chen_2006,Baumann2010,Li2013,PhysRevA.84.053610}. Since the $E_{max}$ is measured in the evolution, it is not a $priori$ clear that it can identify quantum phase transitions. Fig.~\ref{fig:contour}(c) shows the critical curves of the Mott [for $\lvert\left\langle J_{z}\right\rangle / (N/2)\rvert=0$] and the normal [for $\lvert\left\langle J_{z}\right\rangle / (N/2)\rvert=1$] phases. Correspondingly, at the critical point of the quantum phase transition, the maximum stored energy of the CHS QB changes significantly.

\begin{figure}[!ht]
\centering
\includegraphics[width=0.5\textwidth]{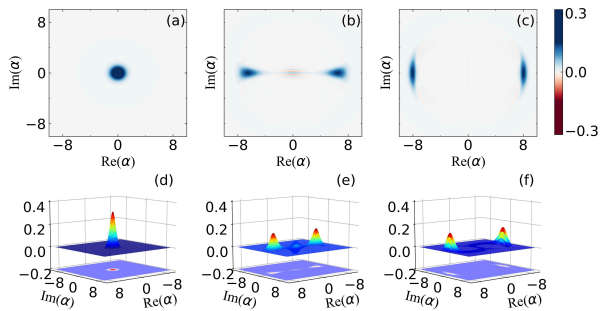}
\caption{\label{fig:wigner}Wigner function near the critical point (the red dots in Fig.~\ref{fig:contour} and Fig. ~\ref{fig:entropy}). The values of $g_{1}$, $g_{2}$ are: (a) and (d): $(-2, 0.2)$, (b) and (e): $(-2, 0.65)$, (c) and (f): $(-2, 1.0)$.}
\end{figure}

We further introduce the Wigner function, which is a way to visualize quantum states using phase-space formalism to describe some physical processes and effects \cite{PhysRev.40.749,HILLERY1984121,Wigner1997,Kim1990,Kohen1997,Querlioz2013,Weinbub2018}. We calculated the ground state cavity Wigner function by randomly selecting three points near the critical point (see Fig.~\ref{fig:contour}) as shown in Fig.~\ref{fig:wigner}. For a fixed spin-spin interaction strength, with the enhancement of cavity-spin coupling, the Wigner function splits from one to two peaks at the critical point, which means that the system shows non-classical properties like coherence and entanglement. However, different from quantum phase transitions, this non-classical phenomenon does not only occur in the region of spin-spin attraction, a similar phenomenon is also observed in the region of spin-spin repulsion.

\begin{figure}[!ht]
\centering
\includegraphics[width=0.35\textwidth]{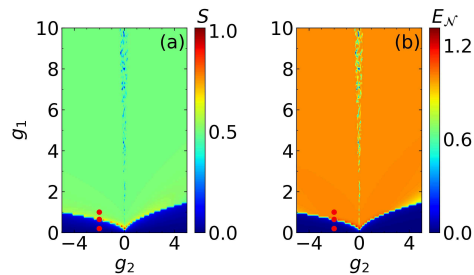}
\caption{\label{fig:entropy}The $S$ and the $E_\mathcal{N}$ as function of the cavity-spin coupling strength $g_{1}$ and the spin-spin interaction strength $g_{2}$ of CHS QB's ground state. The red dots are three randomly selected points at and near the critical point.}
\end{figure}

The analysis of the critical behavior and the non-classical properties of the CHS QB is addressed from the standpoint of the amount of entanglement. We consider the entanglement given by the $von$ $Neumann$ $entropy$ \cite{PhysRevLett.80.2245,PhysRevA.53.2046,RevModPhys.80.517} and the $logarithmic$ $negativity$ \cite{PhysRevLett.95.090503,PhysRevA.65.032314,PhysRevLett.84.4260}. The former is one of the most standard and simple methods to measure entanglement, and the latter is easy to calculate and provides an upper bound on the distillable entanglement. They are defined by
\begin{equation}\label{entropy}
S=- Tr(\rho_{B} \log_{2}\rho_{B}),
\end{equation}
and
\begin{equation}\label{lognega}
E_\mathcal{N}= log_2\Vert\rho^{TB}\Vert_{1},
\end{equation}
where the $\rho_{B} = Tr_{A}(\rho_{AB})$ is the reduced density matrices of subsystems battery part. $\rho^{TB}$ denotes the partial transpose of $\rho$ with respect to the battery part. In Fig.~\ref{fig:entropy}, we illustrate the relation of the $S$ and $E_\mathcal{N}$ concerning the $g_{1}$ and $g_{2}$, showing an obvious critical phenomenon.

For a fixed spin-spin interaction strength, when the cavity-spin coupling strength is less than the critical value, there is no entanglement between the cavity and the spin. Correspondingly, the CHS QB can hardly store energy. As the coupling increases, both the entanglement and the maximum stored energy exhibit critical behavior. The entanglement remains stable after attaining a local maximum at the critical point, but the maximum stored energy continues to increase. Moreover, when there is no interaction between the spins, both energy and entanglement appear discontinuous behavior, which may be caused by the degeneracy of the ground state energy level. Such characteristics indicate that entanglement can be necessary for QB to store more energy, but not sufficient, which is consistent with earlier results \cite{campaioli2018quantum,PhysRevA.101.032115}. Furthermore, our results suggest that some dynamic quantities similar to the maximum stored energy can also carry ground state information such as quantum phase transitions and entanglement.
\section{CONCLUSIONS} \label{section5}
We have introduced the concept of CHS QB, consisting of HS coupling to a single-mode cavity. We have analyzed the influence of parameters such as spin-spin interaction, anisotropy and cavity-spin coupling on QB performance, including the stored energy and the average charging power. Our results demonstrate that the cavity has a positive effect on CHS QB in most cases compared to the HS QB. For fixed spin-spin interactions, the cavity-spin coupling strength can increase the maximum stored energy of the QB, but when there is no interaction between spins, the influence of cavity-spin coupling on the maximum stored energy becomes very weak. The maximum charging power shows a similar pattern to the maximum stored energy, but the maximum charging power is more sensitive to the cavity-spin coupling strength. The effect of the anisotropy on the maximum stored energy and the maximum charging power of the CHS QB depends on the strength of the spin-spin interactions. To be precise, stronger spin-spin interaction increases the maximum stored energy and the maximum charging power of CHS QB but reduces their robustness to anisotropy coefficients. We also investigated the effect of the spin number $N$ on the QB's maximum stored energy and found that it increases linearly with $N$ under different spin-spin interaction strengths. In particular, we have obtained the quantum advantage of the QB's maximum charging power, which approximately satisfies a scaling relation $P_{max} \propto N^{\alpha}$ where scaling exponent $\alpha$ varies with the number $N$ of the spins. Under different spin-spin interaction strengths, for a finite-size system, the CHS QB can lead to a higher average charging power (scaling exponent of the maximum charging power $\alpha=1.5$ and even exceed $1.5$) compared to the HS QB (scaling exponent of the maximum charging power can only reach about $\alpha = 0.75$). Moreover, we find that there is a critical behavior for the maximum stored energy, which is also accompanied by the critical behavior of the quantum phase transitions, Wigner function and entanglement, which demonstrates that entanglement can be a necessary ingredient for QB to store more energy, but not sufficient. The physical quantities that detect quantum phase transitions and entanglement are calculated in the ground state, while the maximum stored energy is calculated in the dynamics. Our study shows that even the dynamics quantities can carry ground state information such as quantum phase transitions and entanglement.
\section*{Acknowledgments}
The work is supported by the National Natural Science Foundation of China (Grant No. 12075193).

\appendix
\section{HS QB}\label{appendix1}
In this Appendix, we first describe the properties of quantum HS QB with magnetic field which we consider as $H_{HS}$. Its ground state serves as the possible initial state of the QB. As shown in Fig.~\ref{fig:model} (a) and (b), the Hamiltonian reads as
\begin{eqnarray}\label{hamHHS}
\begin{split}
H_{HS}=&\omega_{a}\hat{J}_z\\
&+\omega_{a}g_{2}[\hat{J}_{+}\hat{J}_{-}+\hat{J}_{-}\hat{J}_{+}+\gamma(\hat{J}_{+}^{2}+\hat{J}_{-}^{2})\\
&+2 \Delta \hat{J}_{z}^{2}-\frac{N}{2}(2+\Delta)].\\
\end{split}
\end{eqnarray}

\begin{figure}[!ht]
\centering
\includegraphics[width=0.4\textwidth]{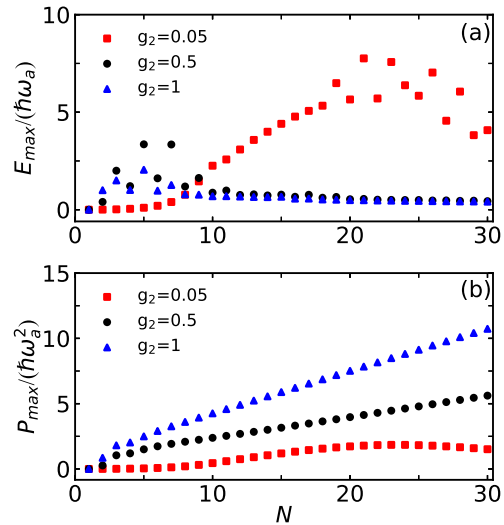}
\caption{\label{fig:variesN1}(a) The maximum stored energy $E_{max}$ (in units of $\hbar$$\omega_{a}$) as a function of $N$ for different values of  $g_{2}$. Results for HS QB refer to $g_{2}$=0.05 (red rectangles), $g_{2} =0.5$ (blue triangles), and  $g_{2} = 1.0$ (black circles). (b) The maximum charging power $P_{max}$ (in units of $\hbar\omega_{a}^{2}$) as a function of $N$. Color coding and labeling are the same as in (a).}
\end{figure}

We further consider the charging process of the extended HS QB in a closed quantum system. Here, the $N$ spins are prepared in ground state $|g\rangle$. A convenient basis set for representing the Hamiltonian is $|j,m\rangle$, where $j(j+1)$ is the eigenvalue of $\hat{J}^2$, and $m$ denotes the eigenvalue of $\hat{J}_z$. With this notation, the HS QB's initial state reads
\begin{equation}\label{HSpsi0}
|\varPsi_{0}\rangle=|N/2,-N/2\rangle.
\end{equation}
\begin{figure}[!ht]
\centering
\includegraphics[width=0.5\textwidth]{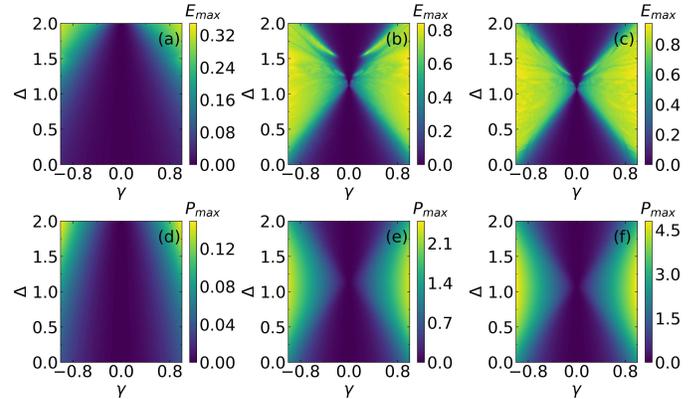}
\caption{\label{fig:deltagamma2}The contour plots of the HS QB's maximum stored energy $E_{max}$ (in units of $\hbar\omega_{a}$) for (a)-(c), and maximum charging power $P_{max}$ (in units of $\hbar\omega_{a}^{2}$) for (d)-(f) with different values of $g_{2}$. The values of $g_{2}$ are: (a) and (d): $g_{2} = 0.05$, (b) and (e): $g_{2} = 0.5$, (c) and (f): $g_{2} = 1$, respectively. Here, we consider the ranges $\gamma\in[-1,1]$, $\Delta\in[0,2]$.}
\end{figure}

The matrix elements of the $H_{HS}$ can be evaluated over the basis set $|j,m\rangle$ using the following relations for ladder operator \cite{Miguel2011,PhysRevA.85.053831,PhysRevE.67.066203}
\begin{equation}\label{Jzhengfu}
\hat{J}_{\pm}|j,m\rangle = \sqrt{j(j+1)-m(m\pm1)}|j,m\pm1\rangle.
\end{equation}
and the matrix elements of the HS Hamiltonian can be found in Appendix~\ref{appendix2}.

The results of HS QB shown in Fig. ~\ref{fig:variesN1} do not clearly show regularity, but we can calculate that for different values of $g_{2}$, for sufficiently large values of $N$, $P_{max}$ follows the scaling law of Eq.~(\ref{Pmaxalpha}):
\begin{equation}\label{Pmaxalpha0.75}
P_{max} \propto N^{0.75}.
\end{equation}

The HS QB's $E_{max}$ does not shows a correlation with $N$ as $N$ increases, and tends to be chaotic.
Fig.~\ref{fig:deltagamma2} shows the larger $g_{2}$ makes HS QB less dependent on anisotropy parameters and also increases the maximum stored energy and charging power of HS QB.

\section{Matrix elements of the Hamiltonian $H$ and $H_{HS}$} \label{appendix2}
The matrix elements of the Hamiltonian $H$ (or $H_{HS}$) can be conveniently evaluated over the basis set $|n,j,m\rangle$ (or $|j,m\rangle$), where $n$ indicates the number of photons, $j(j+1)$ is the eigenvalue of $\hat{J}^2$ and $m$ denotes the eigenvalue of $\hat{J}_z$. Notice that, due to the conservation of $\hat{J}^2$, one can work in a subspace at fixed $j = N/2$. This leads to
\begin{widetext}
\begin{eqnarray}\label{diagCHS}
\begin{aligned}
&\langle {n^{\prime},\frac{N}{2},\frac{N}{2}-q^{\prime}}|H|n,\frac{N}{2},\frac{N}{2}-q \rangle=\\
&\hbar\omega_{c}\{(n+\frac{N}{2}-q)\delta_{q^{\prime},q}+g_{1}[f^{(1)}_{n,\frac{N}{2},\frac{N}{2}-q}\delta_{n^{\prime},n+1}\delta_{q^{\prime},q+1}+f^{(2)}_{n,\frac{N}{2},\frac{N}{2}-q}\delta_{n^{\prime},n+1}\delta_{q^{\prime},q-1}\\
&+f^{(3)}_{n,\frac{N}{3},\frac{N}{2}-q}\delta_{n^{\prime},n-1}\delta_{q^{\prime},q+1}+f^{(4)}_{n,\frac{N}{3},\frac{N}{2}-q}\delta_{n^{\prime},n-1}\delta_{q^{\prime},q-1}]+g_{2}[f^{(5)}_{n,\frac{N}{2},\frac{N}{2}-q}\delta_{q^{\prime},q}+f^{(6)}_{n,\frac{N}{2},\frac{N}{2}-q}\delta_{q^{\prime},q}\\
&+\gamma(f^{(7)}_{n,\frac{N}{2},\frac{N}{2}-q}\delta_{q^{\prime}-2,q}+f^{(8)}_{n,\frac{N}{2},\frac{N}{2}-q}\delta_{q^{\prime}+2,q})+2\Delta(\frac{N}{2}-q)^{2}\delta_{q^{\prime},q}-\frac{N}{2}(2+\Delta)]\}.
\end{aligned}
\end{eqnarray}
\end{widetext}
\begin{widetext}
with
\end{widetext}
\begin{widetext}
\begin{eqnarray}\label{CHSfun}
\begin{aligned}
&f^{(1)}_{k,j,m}=\sqrt{(k+1)[j(j+1)-m(m-1)]},\\
&f^{(2)}_{k,j,m}=\sqrt{(k+1)[j(j+1)-m(m+1)]},\\
&f^{(3)}_{k,j,m}=\sqrt{k[j(j+1)-m(m-1)]},\\
&f^{(4)}_{k,j,m}=\sqrt{k[j(j+1)-m(m+1)]},\\
&f^{(5)}_{k,j,m}=j(j+1)-m(m-1),\\
&f^{(6)}_{k,j,m}=j(j+1)-m(m+1),\\
&f^{(7)}_{k,j,m}=\sqrt{[j(j+1)-(m+1)(m+2)][j(j+1)-m(m+1)]},\\
&f^{(8)}_{k,j,m}=\sqrt{[j(j+1)-(m-1)(m-2)][j(j+1)-m(m-1)]}.
\end{aligned}
\end{eqnarray}
\end{widetext}
\begin{widetext}
and
\end{widetext}
\begin{widetext}
\begin{eqnarray}\label{diagHS}
\begin{aligned}
&\langle{\frac{N}{2},\frac{N}{2}-q^{\prime}}|H_{HS}|\frac{N}{2},\frac{N}{2}-q\rangle=\\
&\hbar\omega_{a}\{(\frac{N}{2}-q)\delta_{q^{\prime},q}+g_{2}[f^{(1)}_{\frac{N}{2},\frac{N}{2}-q}\delta_{q^{\prime},q}+f^{(2)}_{\frac{N}{2},\frac{N}{2}-q}\delta_{q^{\prime},q}\\
&+\gamma(f^{(3)}_{\frac{N}{2},\frac{N}{2}-q}\delta_{q^{\prime}-2,q}+f^{(4)}_{\frac{N}{2},\frac{N}{2}-q}\delta_{q^{\prime}+2,q})+2\Delta(\frac{N}{2}-q)^{2}\delta_{q^{\prime},q}-\frac{N}{2}(2+\Delta)]\}.
\end{aligned}
\end{eqnarray}
\end{widetext}
\begin{widetext}
with
\end{widetext}
\begin{widetext}
\begin{eqnarray}\label{HSfun}
\begin{aligned}
&f^{(1)}_{j,m}=j(j+1)-m(m-1),\\
&f^{(2)}_{j,m}=j(j+1)-m(m+1),\\
&f^{(3)}_{j,m}=\sqrt{[j(j+1)-(m+1)(m+2)][j(j+1)-m(m+1)]},\\
&f^{(4)}_{j,m}=\sqrt{[j(j+1)-(m-1)(m-2)][j(j+1)-m(m-1)]}.
\end{aligned}
\end{eqnarray}
\end{widetext}

\bibliography{reference}

\end{document}